\documentclass[twocolumn,aps]{revtex4}
\usepackage{amssymb}
\usepackage{amsfonts}
\usepackage{amsmath}
\usepackage{graphicx}
\usepackage{xcolor}

\begin{document}

\title{Critical properties of the optical field localization in a
three-dimensional percolating system: Theory and experiment}
\author{Gennadiy Burlak$^{1\ast }$, A. D\'iaz-de-Anda$^{2}$, Boris A. Malomed%
$^{3,4}$, E. Martinez-S\'{a}nchez${^{5}}$, G. Medina-\'{A}ngel${^{1}}$, R.
Morales-Nava$^{6}$, J. J. Mart\'{\i}nez-Ocampo$^{1}$, M.E. de-Anda-Reyes$%
^{2} $, and A. Romero-L\'{o}pez$^{2}$}
\affiliation{${^1}$CIICAp, Universidad Aut\'onoma del Estado de Morelos, Av. Universidad
1001, Cuernavaca, Morelos 62210, M\'exico, E-mail: gburlak@uaem.mx}
\affiliation{${^2}$ Instituto de F\'isica, Benem\'erita Universidad Aut\'{o}noma de
Puebla, P.O. Box J-48, 72570 Puebla, Puebla, M\'{e}xico}
\affiliation{$^{3}$Department of Physical Electronics, School of Electrical Engineering,
Faculty of Engineering, and Center for Light-Matter Interaction, Tel Aviv
University, P.O.B. 39040, Tel Aviv, Israel}
\affiliation{$^{4}$Instituto de Alta Investigaci\'{o}n, Universidad de Tarapac\'{a},
Casilla 7D, Arica, Chile}
\affiliation{${^{5}}$Facultad de Ingenier\'{\i}a, Universidad Aut\'{o}noma de Coahuila,
Blvd.\ Fundadores Km.\ 13 Ciudad Universitaria, Arteaga, Coahuila 25354, M%
\'{e}xico}
\affiliation{${^6}$ Tecnologico de Monterrey, School of Engineering and Sciences, Campus
Puebla, V\'{\i}a Atlixc\'{a}yotl 2301, C.P. 72453, Puebla, Puebla, M\'{e}xico}

\begin{abstract}
We systematically study the optical field localization in an active
three-dimensional (3D) disordered percolating system with light nanoemitters
incorporated in percolating clusters. An essential feature of such a hybrid
medium is that the clusters are combined into a fractal radiation pattern,
in which light is simultaneously emitted and scattered by the disordered
structures. Theoretical considerations, based on systematic 3D simulations,
reveal nontrivial dynamics in the form of propagation of localized field
bunches in the percolating material. We obtain the length of the field
localization and dynamical properties of such states as functions of the
occupation probability of the disordered clusters. A transition between the
dynamical states and narrow point-like fields pinned to the emitters is
found. The theoretical analysis of the fractal field properties is followed by an experimental study of the
light generation by nanoemitters incorporated in the percolating clusters.
The experimental results corroborate theoretical predictions.
\end{abstract}

\maketitle


\section{Introduction}

Disordered photonic materials can transmit and trap light through random
multiple scattering that leads to formation of electromagnetic modes
depending on the structural correlations, scattering strength, and
dimensionality of the system 
\cite{Riboli:2014, Vynck:2012, Flach:2009,Sheng:2010,Wang:2011, Jendrzejewski:2012, Segev:2013a, Wiersma:2013a,Matis:2022,Skipetrov:2018,Cobus:2018}. 
The
Anderson localization was predicted as a non-interacting linear interference
effect \cite{Anderson:1958}. However, in real systems non-negligible
interaction between light and medium takes place. Therefore, an important
aspect of the optical localization is the interplay between nonlinear
self-interaction of light in the disordered medium and the linear Anderson
effect in the same medium \cite{Segev:2013a}. The linear localization of
classical waves can be interpreted as a result of interference between their
amplitudes associated with scattering paths of wave packets propagating
among the diffusers. The study of field transitions in three-dimensional
(3D) disordered optical systems has not yet led to fully established
conclusions, in spite of considerable efforts. The localization transition
may be difficult to reach for the light waves due to the action of various
detrimental effects in dense disordered media required to achieve strong
scattering (see Ref. \cite{Skipetrov:2016a} and references therein). The
experimental observation of optical Anderson localization \cite{Faez:2009a}
just below the field transition in a 3D disordered medium shows strong
fluctuations of the wave function that leads to nontrivial length-scale
dependence of the intensity distribution (multifractality). To study the
field localization in disordered systems, one randomly perturbs parameters
(for instance, the refractive index) around a mean value, and then
identifies guided modes in such a disordered medium, see, e.g., Ref. \cite%
{karbasi:2012} and references therein. It is often assumed that materials
have well-defined mean values of local coefficients, that do not lead to
critical behavior.

Other physically relevant random media are percolating systems, where the
disorder is defined by a value of population probability $p$. A fundamental
feature of such a system is that, in a vicinity of the critical (threshold)
value, $p=p_{\mathrm{cr}}$, the percolation phase transition occurs. It
leads to the formation of long-range connectivity on a scale determined by
the system's length or the size of a spreading percolating cluster. The
spreading cluster has a disordered shape, that essentially affects
properties of the medium \cite{Stauffer:2003}. The resulting disordered
structure is characterized by a non-integer fractal dimension $D_{F}$.

In this work, we aim to study the localization of an optical field in the 3D
percolating disorder, with percolating clusters filled by light nanoemitters
in the excited state. The peculiarity of the setup is that in such materials
the clusters create a fractal nonlinear radiating system where light is both
emitted and scattered by inhomogeneity of the clusters. Such a system may be
considered as a 3D extension of the localization mechanism acting in a
one-dimensional waveguide in the presence of controlled disorder \cite%
{Billy:2008a}. In 3D disordered percolating systems, the optical transport
was observed \cite{Burlak:2009a}, and it was also found that random lasing,
assisted by nanoemitters incorporated in such a disordered structure, can
occur \cite{Burlak:2015}.

The linear theory of the field localization in such a system, that is valid
for very small times (smaller than the laser generation time) was developed
in Ref. \cite{Burlak:2017}. However the consideration of small times does
not make it possible to study nonlinear dynamical effects that emerge on a
longer time scale, close to the lasing threshold in such a system. Then, the
following issues are relevant in this context: (i) the onset of the
optical-field localization in a 3D active time-irreversible nonlinear
system, cf. Ref. \cite{Sheng:2010}, and (ii) the existence of a transition
separating localized propagating optical states and static fields pinned to
random radiating emitters.

The rest of the paper is organized as follows. In Section II we formulate
the model. Section III reports numerical results concerning the lasing by
nanoemitters in the percolating medium. In Section IV we discuss the optical
localization, while in Section V we study the transition separating the
propagating states and narrow field patterns pinned to the emitters. In
Section VI we report results of our experimental investigation of the light
generation by nanoemitters incorporated in the percolating cluster. The last
Section concludes the paper.

\section{Basic equations}

\begin{figure}[tbp]
\centering
\includegraphics[width=0.5\textwidth]
{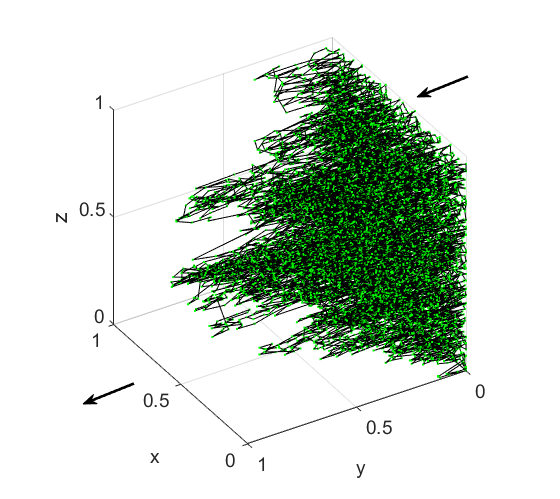}
\caption{(Color online) A typical spatial structure of the incipient
percolating cluster near the percolation threshold at $p=0.3$ in the cube of
size $l_{0}{\times }l_{0}{\times }l_{0}$, with $l_{0}=10^{-4}~m$. To display
necessary details, a numerical grid of size $100{\times }100{\times }100$ is
used in this figure. Only spanning clusters are shown, while internal ones
detached from the entry plane are not displayed. In this configuration, a
considerable number of emitters are incorporated close to the entry
(indicated by the incoming arrow). Solid lines (green color) connect nodes
linked by the variational Fermat's principle, see details in Ref.
\protect\cite{Burlak:2015}. }
\label{Fig1}
\end{figure}

The percolation normally refers to the leakage of fluid or gas through
porous materials. We consider the setting in which percolating clusters are
filled by nanoemitters in an excited state. A typical incipient percolation
cluster has a dendrite shape, see Fig. \ref{Fig1}. In Fig. %
\ref{Fig1} we observe that such a system exhibits two
essentially different domains. The first one (close to the entrance shown by
the arrow) features strong concentration of percolation clusters, which may
lead to enhancement of nonlinear field effects and commencement of lasing.
In the other domain, the concentration is small, hence the local disorder
strength of the medium is small too.

In what follows we address the lasing by nanoemitters incorporated in a
spanning percolating cluster that occupies a large portion of the working
space. First it is instructive to display the 3D spatial structure of the
incipient cluster for the population probability $p$ taking values nearly
the percolation transition. Figure \ref{Fig1} exhibits
such a cluster (all internal disconnected clusters are omitted) for $0.3\leq
p\leq p_{\mathrm{cr}}\approx 0.318$. The figure displays a fractal structure
of the incipient cluster, similar to that produced in Ref. \cite{Burlak:2015}%
. The cluster features a dendrite shape that depends on the actual random
sampling. Re-running simulations with another random seed will lead to a
percolation cluster with a different dendrite structure. \

\section{Lasing by nanoemitters}

As mentioned above, we assume that the percolating clusters with the
dendrite shape are filled by light nanoemitters in the excited state. The
Maxwell's equations for electric $\mathbf{E=E(r,t)}$\ and magnetic $\mathbf{%
H=H(r,t)}$\ fields in the system are%
\begin{equation}
\nabla \times \mathbf{E}=-\mu _{0}\frac{\partial \mathbf{H}}{\partial t}%
\text{, }\nabla \times \mathbf{H}=\varepsilon \frac{\partial \mathbf{E}}{%
\partial t}+\sum_{k}\mathbf{j}(\mathbf{R}_{k},t)\delta _{\mathbf{rR}_{k}}%
\text{\textbf{, } }  \label{MaxwellEqs}
\end{equation}%
where $\mathbf{j}(\mathbf{R}_{k},t)$ ($k=1,..,M$) is the electric current in
radiating emitters set at positions $\mathbf{R}_{k}$. The equation for the
polarization density $\mathbf{P}_{k}\mathbf{\equiv P(\mathbf{R}_{k},}t%
\mathbf{)}$ in the cluster filled by the emitters (four-level atoms), with
level occupations $N_{m,k}$ ($m=0,1,2,3$ is the level's number),\ is%
\begin{equation}
\frac{\partial ^{2}\mathbf{P}_{k}}{\partial t^{2}}+\Delta \omega _{a}\frac{%
\partial \mathbf{P}_{k}}{\partial t}+{\omega _{a}^{2}}\mathbf{P}_{k}=\frac{%
6\pi \varepsilon _{0}c^{3}}{\tau _{21}\omega _{a}^{2}}(N_{1,k}-N_{2,k})%
\mathbf{E}_{k}.\text{ }  \label{PolarizEq}
\end{equation}%
To complete the model we add the rate equations for the level occupations of
emitters, $N_{i,k}=N_{i}(\mathbf{\mathbf{R}_{k}},t)$:%
\begin{equation}
\frac{\partial N_{0,k}}{\partial t}=-A_{r}N_{0,k}+\frac{N_{1,k}}{\tau _{13}}%
\text{, }\frac{\partial N_{3,k}}{\partial t}=A_{r}N_{0,k}-\frac{N_{3,k}}{%
\tau _{02}},  \label{EqForN_03}
\end{equation}%
\begin{equation}
\frac{\partial N_{1,k}}{\partial t}=\frac{N_{2,k}(t)}{\tau _{(21),(32)}}-%
\frac{(\mathbf{j}\cdot \mathbf{E})_{k}}{\hbar \omega _{a}}-\frac{N_{1,k}}{%
\tau _{13}}\text{, }\frac{\partial N_{2,k}}{\partial t}=\frac{N_{1,k}(t)}{%
\tau _{12}}+\frac{(\mathbf{j}\cdot \mathbf{E})_{k}}{\hbar \omega _{a}}-\frac{%
N_{2,k}}{\tau _{02}}\text{,}  \label{EqForN_12}
\end{equation}%
Here $\Delta \omega _{a}=\tau _{21}^{-1}+2T_{2}^{-1}$, where $T_{2}$ is the
mean time between dephasing events, $\tau _{21}$ is the time of the
spontaneous decay from the second atomic level to the first one, and $\omega
_{a}$ is the radiation frequency (see, e.g., \cite{Siegman:1986}). Electric
field $\mathbf{E(r,}t\mathbf{)}$ and current $\mathbf{j(\mathbf{R}_{k},}t%
\mathbf{)}=\partial \mathbf{P(\mathbf{R}_{k},}t\mathbf{)}/\partial t$ are
found from the Maxwell's equations, together with equations for densities $%
N_{m}\mathbf{(\mathbf{R}_{k},}t\mathbf{)}$ of atoms residing at the $m$-th
level. An external source excites the emitters from the ground level ($m=0$)
to the third level at certain rate $A_{r}$, which is proportional to the
pumping intensity \cite{Soukoulis:2000}. After a short lifetime $\tau _{32}$%
, the emitters transfer nonradiatively to the second level. The second and
first levels are the top and bottom lasing levels, respectively. Emitters
decay from the top level to the bottom one through spontaneous or stimulated
emission, $(\mathbf{j}\cdot \mathbf{E})/\hbar \omega _{a}$ being the
stimulated-radiation rate. Finally, emitters can also decay nonradiatively
from the first level to the ground level. The lifetimes and energies of the
top and bottom lasing levels are $\tau _{21}$, $E_{2}$ and $\tau _{10}$, $%
E_{1}$, respectively. The individual radiation frequency of each emitter is
then $\omega _{a}=\left( E_{2}-E_{1}\right) /\hbar $.

To study the dynamics, we solved the semiclassical equations (\ref%
{MaxwellEqs}) - (\ref{EqForN_12}) that couple field $\mathbf{E(r,t)}$,
polarization density $\mathbf{P(\mathbf{R}_{k},}t\mathbf{)}$, and
occupations of the emitters' levels $N_{i}\mathbf{(\mathbf{R}_{k},}t\mathbf{)%
}$. In the following Section we use the numerical finite-difference
time-domain algorithm \cite{FDTD} implemented on the numerical grid of size $%
100\times 100\times 100$ to produce 3D solutions to Eqs. (\ref{MaxwellEqs})-(%
\ref{EqForN_12}). The solutions are presented in terms of scaled time $t$
and coordinates $\mathbf{r}$ defined by $t\rightarrow {tc}_{0}{/l_{0}}$ and $%
\mathbf{r}\rightarrow \mathbf{r}{/l_{0}}$, where
\begin{equation}
l_{0}=23.0~\mathrm{\mu }\text{\textrm{m}}  \label{l_0}
\end{equation}%
is a typical spatial scale (corresponding to our experimental setup, see
below), and $c_{0}$ is the light speed in vacuum.

We aim to calculate the time-averaged integral emission of electromagnetic
energy from a cubic sample $(x,y,z)\in \lbrack 0,l_{0}]$. The total output
flux of energy can be written as
\begin{equation}
I=\oint_{S}(\mathbf{K}\cdot \mathbf{n})dS=I_{x}+I_{y}+I_{z},
\label{balance1}
\end{equation}%
where $\mathbf{K}$ is the time averaged Pointing vector, $\mathbf{n}$ is the
normal unit vector on surface $S$ of cube, and $I_{x,y,z}$ are energy fluxes
(intensities) emitted from both faces of the cube perpendicular to a
particular direction. To calculate the energy flux defined by Eq. (\ref%
{balance1}), we solved numerically the equations that couple polarization
density $\mathbf{P}$, electric field $\mathbf{E}$, and four-level
occupations of the emitters.

\begin{figure}[tbp]
\centering
\includegraphics[width=0.5\textwidth]
{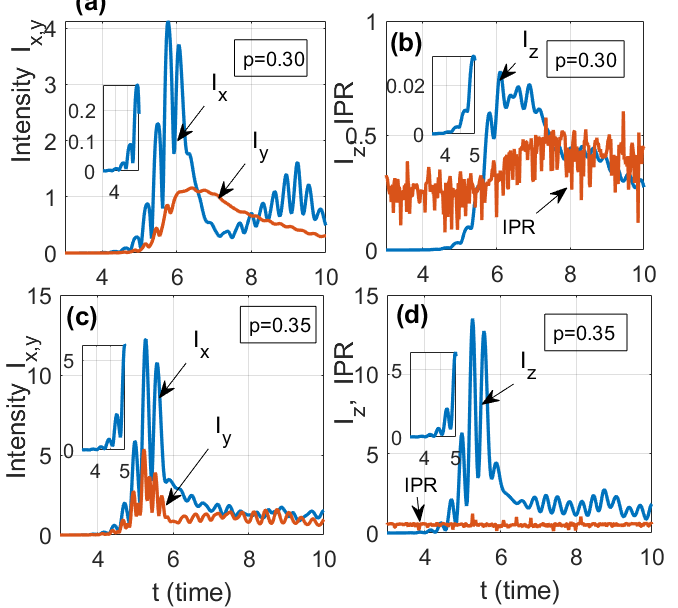}%
\caption{(Color online) The lasing output (stimulated by the nanoemitters
embedded in the percolating clusters) with intensities \textcolor{black}{(arb.units)} (a) $I_{x,y}$, (b) $%
I_{z}$ [defined by Eq. (\protect\ref{balance1})], and the inverse
participation ratio (IPR), defined as per Eq. (\protect\ref{InvPartRat}), as
functions of time $t$. In (a,b) the percolating probability is $p=0.30$,
while in (c,b) it is $p=0.35$. Insets in panels (a)-(d) display the initial
evolution of intensities $I_{x},I_{z}$ at small times $t<5$ (for respective
values $p$), from which one observes that lasing starts at time $%
t_{C}\approx 4.1$, and then exponentially rapidly approaches the peak values
at $t\approx 4.8$. After that, the nonlinear dynamics commences. }
\label{Fig2}
\end{figure}

Figure \ref{Fig2} shows the evolution of the field energy fluxes
$I_{x,y,z}$ on the output surface of the sample. In panel \ref%
{Fig2}(a) we observe that initial field amplitudes are very
small, so that the system of Eqs. (\ref{MaxwellEqs})-(\ref{EqForN_12}) is
effectively linear at the initial stage of the evolution.

\begin{figure}[tbp]
\centering
\includegraphics[width=0.5\textwidth]
{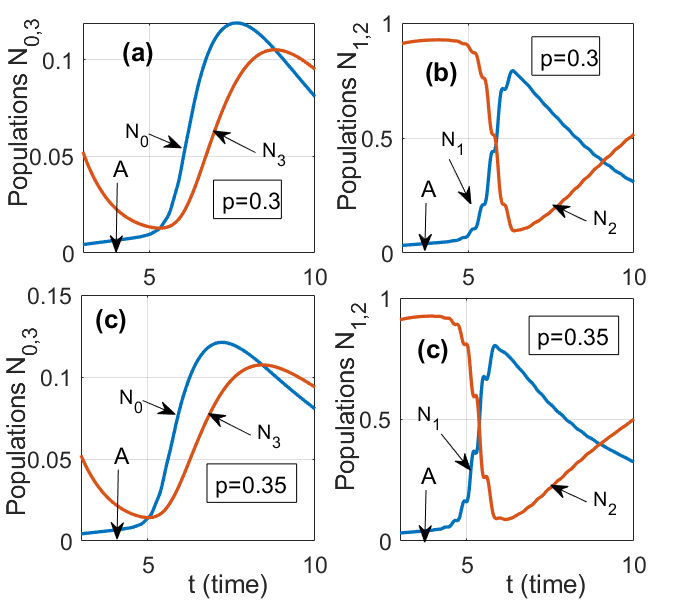}
\caption{(Color online) The evolution of populations $N_{j}$ ($j=0,1,2,3$)
of the four levels in the radiating nanoemitters incorporated in the
percolating clusters: (a) $N_{0,3}$ and (b) $N_{1,2}$ vs. time $t$. In (a,b)
the percolating probability is $p=0.30$, and in (c,b) it is $p=0.35$.
Vertical arrows $\mathrm{A}$ indicate the critical time, $t_{C}\approx 4.1$,
at which the lasing starts, cf. Fig. \protect\ref{Fig2}. The
nonlinear dynamics commences at $t>t_{C}$. }
\label{Fig3}
\end{figure}

At times $t<t_{C}\approx 4.1$ synchronization of initially random phases of
the emitters in the clusters occurs. Then, at times close to $t=t_{C}$, Fig. %
\ref{Fig2} demonstrates the generation of a strong field output
that corresponds to the onset of lasing in the ensemble of excited emitters
embedded in the clusters. Insets in Figs. \ref{Fig2}(a-d)
display the initial dynamics of intensities $I_{x,y}$ and $I_{z}$ at $t<5$
(for the respective values of the percolation probability, $p=0.30$ and $%
0.35 $), showing that the lasing starts at time $t_{C}\approx 4.1$ and then
it exponentially rapidly attains peak values at time $t\approx 4.8$, at
which the nonlinear dynamics commences. Starting from $t=t_{C}$, the
disordered percolating system features the transition to the optical
generation, when, in addition to the amplification, the optical waves
undergo multiple scattering in the percolating fractal medium. As the lasing
starts in all field modes, the optical field and radiating nanoemitters form
a strongly coupled nonlinear dynamical system. To evaluate the strength of
the field distribution, we calculated the inverse participation ratio (IPR)
of the field, defined as%
\begin{equation}
IPR=\frac{l_{0}^{3}\int \left\vert \mathbf{E}\right\vert ^{4}d^{3}\mathbf{r}}{%
\left( \int \left\vert \mathbf{E}\right\vert ^{2}d^{3}\mathbf{r}\right) ^{2}}%
,  \label{InvPartRat}
\end{equation}%
where $l_{0}$ is the sample's size, and the integration is performed over
the entire system. Figure \ref{Fig2}(b) demonstrates complex
oscillatory behavior of IPR, which implies a high degree of randomicity in
the system. However, after the onset of the lasing, IPR as a whole exhibits
a trend to transition towards a more ordered state of the field. The
dynamics of the population of the four levels of the nanoemitters, which
follows the phase synchronization and the onset of the lasing, is displayed
in Fig. \ref{Fig3}. In particular, the evolution of
populations $N_{1,2}$ that leads to the start of the lasing is shown in Fig. %
\ref{Fig3}(b).

\section{The localization}

Random multiple scattering of optical waves leads to field localization, in
the form of well-trapped modes. Such states are defined by a characteristic
average localization radius of the optical field. If the field amplitude in
the localized mode is small, it propagates as a free one in the linear
regime.

As this phenomenon occurs in the highly disordered system, it may be
identified as a variant of the 3D optical-field localization. Similar to the
widely adopted approach in the field localization theory, see Ref. \cite%
{Genack:2010} and references therein, we define the dimensionless
localization measure, $g$, as the ratio of two spatial scales: the mean free
paths of photons in a sample, $\bar{l}$, and wavelength $\lambda $ of the
radiation of emitters embedded in the percolating cluster. Length $\bar{l}$
is proportional to the (dimensionless) distance $A_{\mathrm{3D}}$ which is a
number of nodes in the numerical grid, which a photon can travel in the 3D
sample, passing $N_{e}(p)$ emitters without visiting the same region twice. $%
A_{\mathrm{3D}}$ is calculated directly in our simulations by dint of the
TSP (Traveling Salesman Problem) technique \cite{Press:2002}. Accordingly,
the dimensionless average free path is $A_{\mathrm{3D}}/N_{e}(p)$, its
dimensional counterpart being $\bar{l}=l_{0}A_{\mathrm{3D}}/N_{e}$, where $%
l_{0}$ is the basic length scale (\ref{l_0}) defined above. Thus, the
localization measure is

\begin{equation}
g(p)=\frac{l_{0}}{\lambda }\frac{A_{3N}}{N_{e}(p)}.  \label{param g}
\end{equation}

For the percolating grid considered here, the numerically computed
dependence $g(p)$ is presented in Fig. \ref%
{Fig4} for typical values of physical
parameters: $l_{0}=23~\mathrm{\mu }${m}, $\lambda =0.5~\mathrm{\mu }$m.

\begin{figure}[tbp]
\centering
\includegraphics[width=0.5\textwidth]
{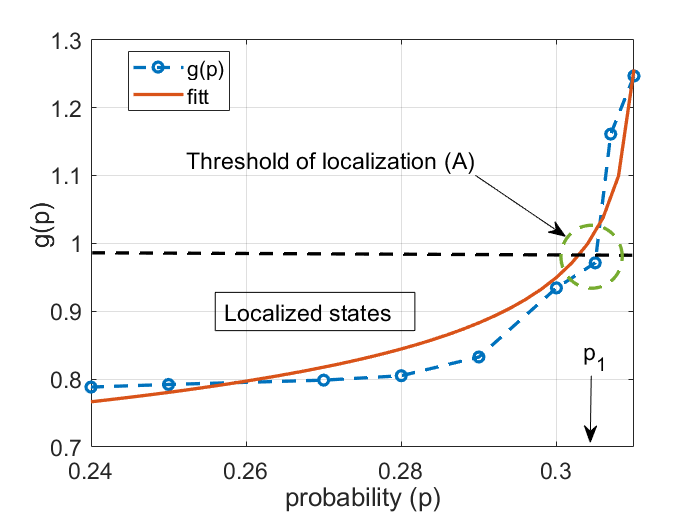}
\caption{(Color online.) Dependence $g(p)$, as produced by numerical
solution of Eq. (\protect\ref{param g}) (blue dash line) for typical values
of physical parameters: field wavelength $\protect\lambda =0.5~\mathrm{%
\protect\mu }$m and scale length $l_{0}=23~\mathrm{\protect\mu }$m. The red
line is the analytical fitting by the probe function $g_{f}(p)=b+|p-a|^{m}$
that gives $a=0.31,m=-0.074$, which turns out to be close to critical
percolation $p_{c}$ and critical exponent $\protect\nu $ respectively, and $%
b=-0.45$. The Ioffe-Regel localization condition, $g\approx 1$, is fulfilled
closely of the percolation threshold, $p\approx p_{c}=0.318$. Large values
of $p$ produce very long free paths $A_{\mathrm{3D}}$, corresponding to $g>1$%
, where the localization does not occur. The region of $g(p)<1$ is indicated
as one maintaining localized states. Arrow A indicates the area of the
localization threshold, which is $3.01<p<3.1$. The exact value of
probability $p$ corresponding to the point $g(p)=1$ is calculated from $%
g_{f}(p)=1$, that gives $p_{1}=0.304$. It is difficult to obtain more
precise values because of the enhacement of fluctuations in the area of the
percolation phase transition at $p\approx p_{c}$ \protect\cite{Stauffer:2003}.}
\label{Fig4}
\end{figure}

The Ioffe-Regel localization condition, $g=1$ \cite{Ioffe,Sheng:2010}, takes
place close to the percolation threshold, $p$ $\approx 0.318$, indicated by
arrow (A) in Fig. \ref{Fig4}. Further increase
of $p$ leads to larger values of the above-mentioned dimensionless distance $%
A_{\mathrm{3D}}$, hence the localization does not occur at $g>1$, while it
persists at $g(p)<1$. It is difficult to calculate a precise value of
probability $p$ corresponding to the point $g(p)=1$ because of significant
increase of fluctuation in the area of the percolation phase transition at $%
p\approx p_{c}$ \cite{Stauffer:2003}.

\begin{figure}[tbp]
\centering
\includegraphics[width=0.5\textwidth]
{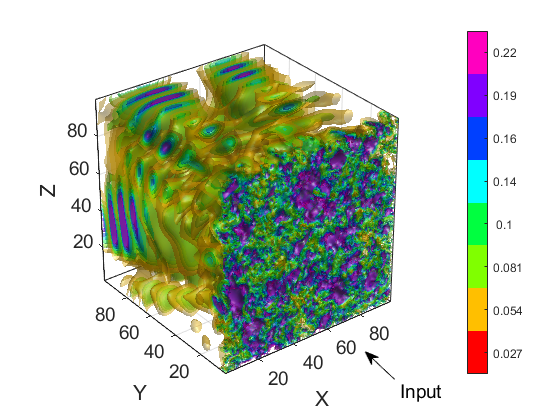}
\caption{(Color online.) The 3D isosurface of the emitted field at time $%
t=11 $ for $p=30$. The fractality of the field gives rise to the complex
structure with \textquotedblleft wormholes" and field bunches.}
\label{Fig5}
\end{figure}

Generally, the structure of the radiated field in the 3D percolating medium
is quite intricate, see Fig. \ref{Fig5},
which displays the isosurface of the 3D emitted field at time $t=11$. We
observe an indented shape of light structures with bunches and wormholes due
to the fractality of the radiating system.
\begin{figure}[tbp]
\centering
\includegraphics[width=0.5\textwidth]
{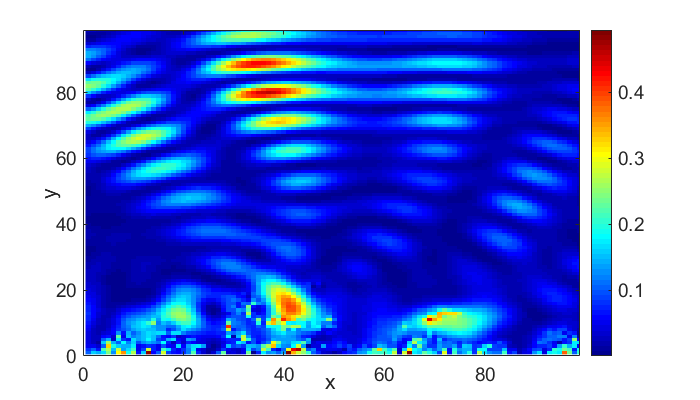}
\caption{(Color online) The normalized field \textcolor{black}{(arb.units)} in the midplane $\left(
x,y\right) $ of the 3D system, corresponding to Fig. \ref{Fig5} at $z=L/2$ for $p=0.3$. In the input  \textcolor{black}{area (small $y$)},
only narrow point-like fields generated by static random emitters are
observed. Farther, field bunches with a smooth localized shape and large
amplitudes arise beyond the emitter's peaks. Thus, lasing in the percolating
disordered system stimulates the localization of the emitted field.}
\label{Fig6}
\end{figure}

Figure \ref{Fig5} demonstrates the
screenshot of the system after some transition time $t=11$ (see the field
peaks in Fig. \ref{EqForN_03}), when dynamics becomes nonlinear. The field
peaks detach from the emitters and generate 3D field bunches in the course
of the evolution. In what follows below, we address the dynamics and
structure of the localized fields confined in the light bunches. A
quasi-spherical field bunch is referred to as a bounded 3D domain,
schematically separated into concentric spherical layers numbered from the
center to periphery by numbers $j$, with field $\left\vert E_{j}\right\vert $
being a decreasing function of $j$, i.e., $\left\vert E_{0}\right\vert
>\left\vert E_{1}\right\vert >\ldots >\left\vert E_{r}\right\vert >0$, where
$r$ is the number (effectively, radius) of the concentric layers. Each bunch
is characterized by two parameters, viz., radius $r$ and the value of the
field at the center, $\left\vert E_{0}\right\vert =E_{\max }$ (a more
precise consideration of such a 3D localization problem requires the use of
huge computational resources). One of these bunches is shown in Fig. \ref%
{Fig7}, where one can clearly observe the structure of the
localized field with a large amplitude $E_{\max }$ and characteristic radius
$r$ at time $t=11$. We have found that such well-established localized
objects are generated at various locations in the 3D medium due to
disordered positions of the radiating emitters and the fractal shape of the
percolating clusters.
Fig.\ref{Fig6} exhibits the normalized field \textcolor{black}{(arb.units)} in the midplane $\left(
x,y\right) $ of the 3D system, corresponding to Fig. \ref{Fig5} at $z=L/2$ for $p=0.3$. 
\textcolor{black}{
In the region of irregular point-like emitters (small y in Fig.6) the strong (nonlinear) fields  are generated dynamically with reproducing the localized shape of the emitter clusters. This leads to the following: the field pulses bounce off the emitters and then propagate independently in the form of inhomogeneous (smooth) waves. The spatial localization of such pulses (in a transversal plane) can be seen in Fig. 6. The localized bunches propagate from the emitters positions and thereafter arrive the output boundaries, where they are observed as the temporal field peaks, see Fig. 2. Fig. 5 displays the details of such a dynamics in 3D.}

\begin{figure}[tbp]
		\includegraphics[width=0.5\textwidth]
		{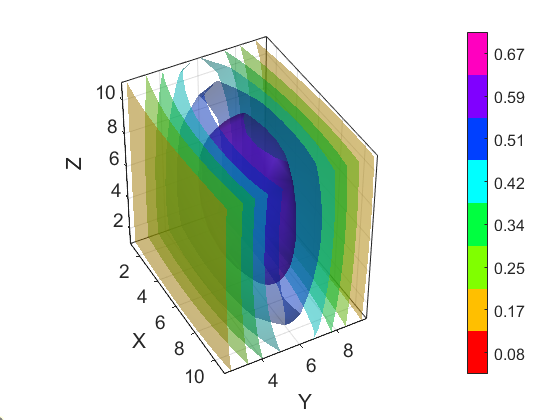}
	\caption{(Color online) The shape of a localized field bunch, with a high
		field amplitude at the center, $E_{\max }=0.98$, and radius $r=5$ at $t=11$,
		for $p=0.3$,. The center of the structure is located at the point with
		coordinates $\left\{ x=14,y=18,z=73\right\} $. In the general case, the
		localized bunches are centered at random positions.}
	\label{Fig7}
\end{figure}

\section{The transition to the localization}

Due to the statistical nature of the field localization, our approach here
is based on multiple realizations of the disordered percolating structure,
sampled over a statistically identical ensemble of the emitters. In active
media (where the time reversibility is broken), such an approach has to
develop the description of the localization strength in terms of the
time-average states. In this case, the percolating population probability of
the host material becomes the control parameter that allows one to
distinguish localized and extended modes. 

Because of the inhomogeneity of
the distribution of the disordered emitters and fractal shape of the
percolating clusters, the radiated field acquires an essentially non-uniform
shape featuring the above-mentioned dynamical bunches. Therefore, in what
follows below, we consider percolation probability $p$ as the parameter
which determines the average radius of the bunches. A region where such a
radius is a persistent feature is identified as the localization domain.
Localization-transition zones are defined as regions where the localization
radius changes sharply. It is worth noting that the velocity of the
dynamical optical bunches can be well established only beyond the area of
the active emitters, as in that area the propagating field is coupled to the
light emitters and thus cannot be separated from the narrow point-like field
of the static nanoemitters.

\begin{figure}[tbp]
\centering
\includegraphics[width=0.5\textwidth]
{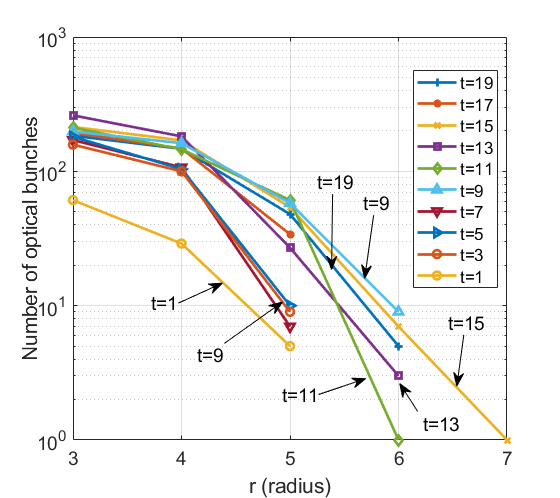} 
\caption{(Color on line) The number of the optical bunches vs. their width
(radius $r$) at different times $t$, for the percolation probability $p=0.3$%
. It is seen that an increasing number of states with larger $r$ are created
in the system in the course of the evolution.}
\label{Fig8}
\end{figure}

In what follows we study the dependence of the number of bunches $M$ at
different times $t$ on radius $r$, for a fixed percolation probability, $%
p=0.3$, which is slightly smaller than the above-mentioned threshold value, $%
p_{\mathrm{cr}}=0.318$ \cite{Stauffer:2003}, \cite{Wang:2013a} Figure \ref%
{Fig8} shows such a dependence for scaled times $t=1,...,19$. It
is observed that, below the lasing threshold ($t\leq 5$, see Fig. \ref%
{Fig6}), there are many field peaks with
small radius $r$. Most of them correspond to narrow point-like field packets
generated by the static randomly distributed light emitters, cf. Fig. \ref%
{Fig6}. However, Fig. \ref{Fig8}
demonstrates that, after the lasing starts (at $t>5$), the number of the
optical bunches with large radius $r$ drastically increases. Besides, Fig. %
\ref{Fig6} shows that such bunches (with
large $r$) are situated mainly beyond the percolation area. Moreover, as the
same figure shows, the bunches feature a well-defined shape; they are no
longer pinned to the static emitters, migrating to the region where emitters
are not present. These features allow one to identify such dynamical bunches
as 3D zones of the nonlinear optical-field localization emerging in the
active percolating clusters with the fractal structure.

While\ Fig. \ref{Fig8}\ displays typical dependences of the
number of localized optical bunches, at different times, on the bunch's
width (radius) for the fixed percolation probability, $p=0.3$. Due to the
statistical nature of the field localization, it is also relevant to
calculate the time average of the dependencies shown in Fig. \ref%
{Fig8}. These numerical results allows fitting to an analytical
approximation with physically meaningful parameters. Accordingly,\ Fig. \ref%
{Fig9} shows a typical dependence the number of
localized optical bunches averaged over the simulation time $t=20$ (blue
asterisks) as a function of the bunch's width (radius $r$) at fixed $p=0.30$.

\begin{figure}[tbp]
\includegraphics[width=0.5\textwidth]
{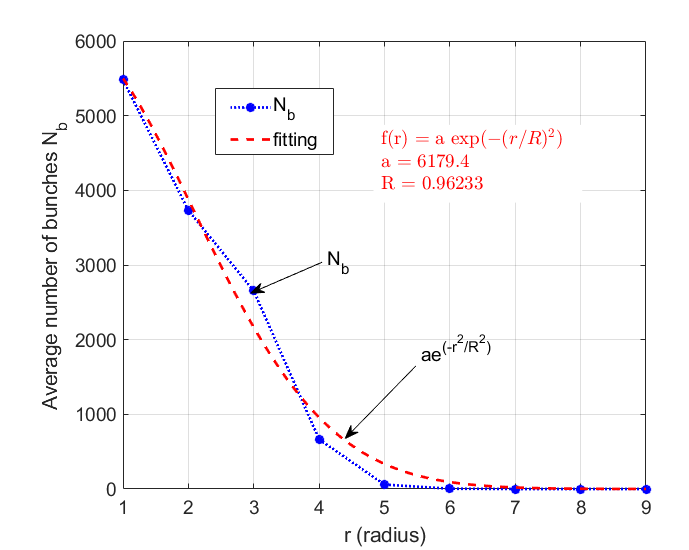} 
\caption{(Color on line) Blue asterisks: the number $N_{b}$ of localized
optical bunches, averaged over $100$ different realizatins of the random
system with $p=0.3$, for simulation time $t=20$, vs. the bunch's radius $r$.
The red dashed line shows the fitting function defined as per Eq. (\protect
\ref{f}), with the fitting parameter $R$ representing the average width of
the bunches. The error of the fitting is less than a few percents. }
\label{Fig9}
\end{figure}

\begin{figure}[tbp]
\includegraphics[width=0.5\textwidth]
{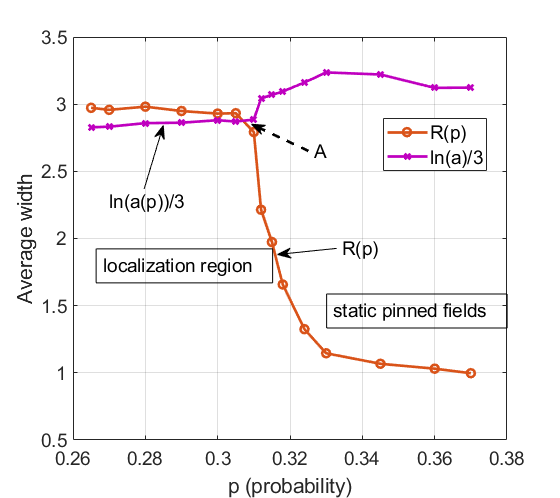} 
\caption{(Color online) The fitting parameters in Eq. (\ref{f}) vs.
the percolation probability $p$. Arrow A indicates the localization
transition that matches to the critical point shown by arrow (A) in Fig.
\protect\ref{Fig4} and the percolating
threshold $p_{c}$.}
\label{Fig10}
\end{figure}

The red dashed line in Fig. \ref{Fig9} shows the fitting
function,
\begin{equation}
f(r)=a\exp (-r^{2}/R^{2}),  \label{f}
\end{equation}%
with fitting parameters $a(p)$ and $R(p)$, where $R(p)$ is the average
radius of the localized bunches at a fixed value of the percolation
probability $p$. Additional computations demonstrates that the same fitting,
with smooth functions $a(p)$ and $R(p)$, is equally accurate for other
relevant values of the percolation probability $p$. The respective
dependences $a(p)$ and $R(p)$ are plotted in Fig. \ref{Fig10}.
One observes in the figure that, close to $p=p_{\mathrm{cr}}\simeq 0.318$,
there occurs a well-defined transition from the average bunch's width $R=3$
to $R=1$. Recall that a narrow field structure with size $r=1$ corresponds
to the point-like static field pinned to the emitter position. Thus, these
results confirm the above conclusion that $p=p_{\mathrm{cr}}$ is the
transition point (the mobility edge) which separates two different areas: at
$p>p_{\mathrm{cr}}$ there exist only static fields pinned to the radiating
nanoemitters, while at $\ p<p_{\mathrm{cr}}$ the dynamical localized field
bunches emerge, which then migrate to the area free of emitters.
The existence of a similar transition point was deduced in Ref. \cite%
{Burlak:2017} from different considerations, making use of the
above-mentioned Ioffe-Regel criterion, $g=1$ [see Eq. (\ref{param g})] for
the 3D field localization. It is worthy to mention that probability $p\simeq
0.31$ is close to the percolation-transition point $p_{\mathrm{cr}}$\ in the
cubic 3D lattice \cite{Burlak:2017}, cf. Fig. \ref%
{Fig4}.

\begin{figure}[tbp]
\centering
\includegraphics[width=0.5\textwidth]
{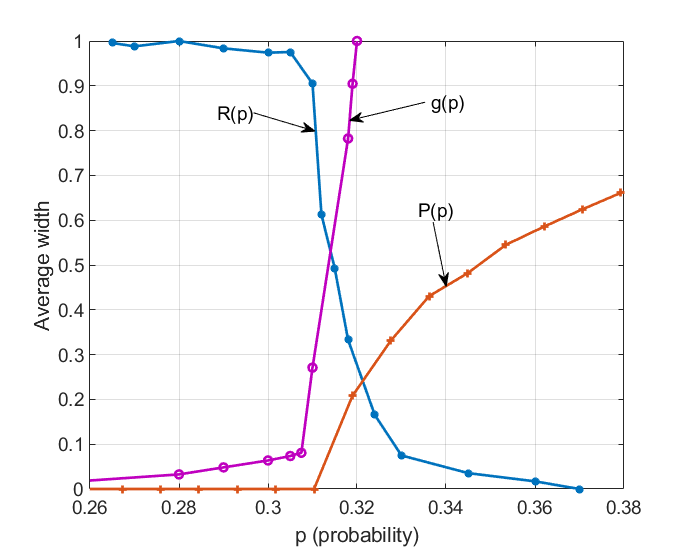}
\caption{(Color online.) The comparison of the optical localization
characteristics close to the percolation threshold $p_{c}=0.318$: the
normalized average radius of the localized optical bunches, $R(p)$, the
generalized conductivity, $g(p)$ [defined as per Eq. (\ref{param g})], 
and the percolation order parameter \cite{Percolation}, $P=P(p)$, vs. the percolation probability, $p$. The transition points produced by
all these dependencies are quite close.}
\label{Fig11}
\end{figure}

Different approaches are used\ above to identify the transition for the
optical field in the percolating materials. It is instructive to compare the
results produced by those approaches near the percolating threshold, $p_{%
\mathrm{cr}}=0.318$, for normalized average radius of the localized optical
bunches, $R(p)$, generalized conductivity $g(p)$ [see Eq. (\ref{param g})],
and the percolation order parameter, $P(p)$, as functions of the percolation
probability $p$ [the order parameter is defined as a probability that a
particular site of the grid approximating the medium belongs to a
percolation cluster piercing the entire sample \cite{Percolation}.
The comparison is produced in Fig. \ref{Fig10}, which shows that
the points of the localization transition, predicted by these different
characteristics, are close to being identical.

\section{Experimental studies of the emission spectra}

In this Section we present experimental findings concerning optical
radiation from nanoemitters incorporated in a porous ceramic matrix, which
offers a realization of the setting considered above in the theoretical form.%
Our experimental study is based on the observation that an
important mechanism for the generation of visible light is the up-conversion
(UC) process by which at least two low-energy excitation photons, typically
in the near infrared, are converted into one visible emission photon of
higher energy, see \cite{Solis:2010a}. In this case the strong green and red
visible emissions can be obtained from ZrO$_{2}:$ Yb$^{+3}$ nanocrystals.
The latter nanocrystals with different concentrations were incorporated into
porous ceramics (approximately $1~$mm size) and then deposited on the inner
surface of empty pores. The nanocrystals were prepared in three different
configurations, (i) pure (undoped) ZnO2, (ii) ZnO2 dopped by Yb with $1.23\%$
concentration (hereafter referred to as sample E4), and (iii) ZnO$_{2}$
doped Yb at $2.36\%$ concentration (referred to as sample E5). For optical
measurements, slices with size about $1~\mathrm{\mu }$m are cut from the
bulk material. The optical measurements of absorption and emission spectra
were performed for every sample.

Since the UC emission is strongly influenced by the content of Yb$^{+3}$
ions, the embedded nanoemitters generate random emission, being excited by
an external laser. The experimental setup is outlined in Fig. \ref%
{Fig12} (details of the technology used for synthesizing the
nanoemitters are presented in Appendix).
\begin{figure}[h]
\centering
\includegraphics[width=0.5\textwidth]
{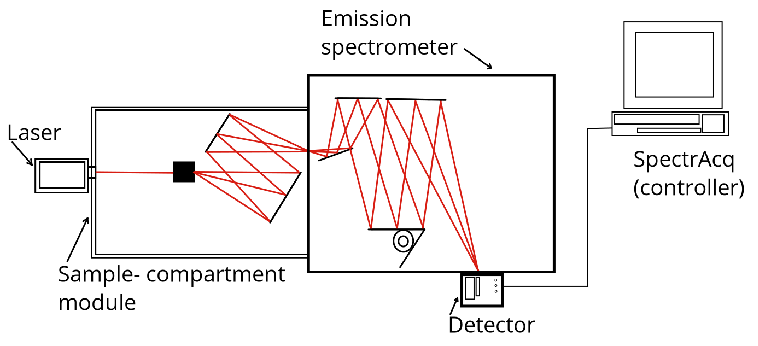}  
\caption{The scheme of the experimental setup built for measuring the
emission spectrum. The small black square designates the sample.}
\label{Fig12}
\end{figure}

We measured the emission spectra of the ceramic samples carrying the
embedded undoped and Yb-doped nanocrystals, with the above-mentioned values
of the dopant concentration, $1.23\%$ and $2.36\%$, respectively, using a
Horiba Jobinyvon NanoLog FR3 spectrofluorometer \cite{Horiba}. The samples
were excited by an external $970$ nm pump. The beam emerging from the sample
was coupled into the emission spectrometer module and finally passed to the
detector. The signal from the detector was sent to the SpectrAcq controller
for data processing.


It is seen that the emission bands for both samples with either value of the
concentration are approximately located between $470$ and $570$ nm. It is
observed too that the emission from the sample with the $2.36\%$ Yb
concentration is $\approx 3$ times more intense than from the one with the $%
1.23\%$ concentration. Thus, it is demonstrated experimentally that the
Yb-based nanocrystals incorporated in the porous ceramics emit in the
green-wavelength range when the ceramic sample is excited by the external
laser with the central wavelength of $970$ nm (see the inset in Fig. \ref%
{Fig13}). Naturally, the radiation intensity increases with the
growth of the concentration of nanocrystals in the ceramic host material.%
\begin{figure}[tbp]
\centering
\includegraphics[width=0.5\textwidth]
{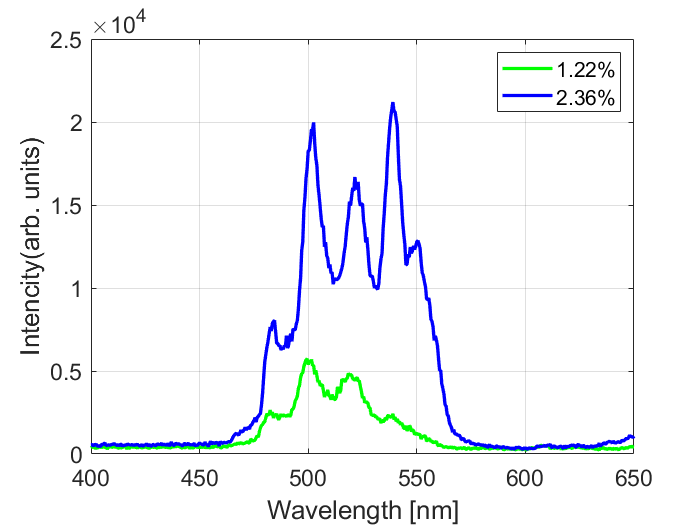}  
\caption{Emission spectra (in arbitrary units) of the Yb-doped ceramics with
different Yb concentrations, \textit{viz}., $1.23\%$ (see details in the
text).}
\label{Fig13}
\end{figure}

\begin{figure*}[tbp]
\centering	
\includegraphics[width=1\textwidth]
{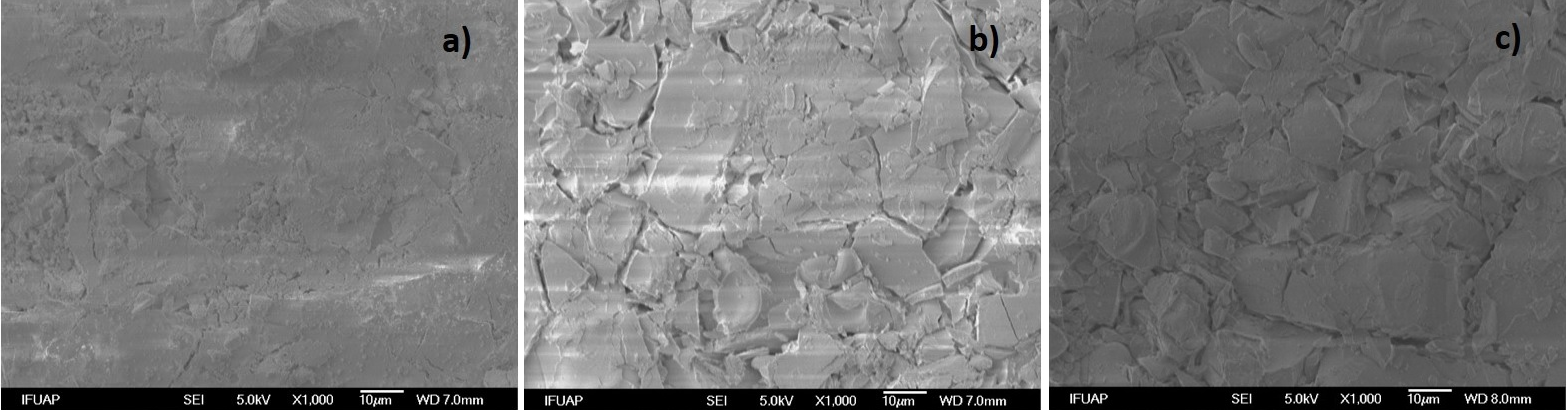} 
\caption{SEM images of a) the undoped ZrO$_{2}$ sample, and doped ones, E4
and E5, with the Yb concentration $1.23$\% [panel b)], and $2.36$\% [panel
c)], respectively.}
\label{Fig14}
\end{figure*}
Figure \ref{Fig14} shows the SEM (scanning-electron-microscope)
image of the surface morphology of the undoped ZrO$_{2}$ sample and Yb-doped
ones (E4 and E5 correspond to the $1.23$\% and $2.33$\% dopant
concentrations, respectively). Figure \ref{Fig14} demonstrates that
the porosity of the samples increases with the increase of the Yb
concentration, in comparison with the undoped sample, which is displayed in
Fig. \ref{Fig14}a). The aggregate sizes of samples E4 and E5,
composed of small crystals, are similar, taking values between $10$ and $30$
$\mathrm{\mu }$m, while more grain boundaries are observed in E5, see Fig. %
\ref{Fig14}c), hence the latter one has a higher level of porosity.
The average pore sizes for the samples are $1$ $\mathrm{\mu }$m, $0.5$ $%
\mathrm{\mu }$m, and $2~\mathrm{\mu }$m for a) undoped ZrO$_{2}$, and doped
samples, viz., E4 in b) and E5 in c), respectively. From the X-ray
diffractograms (XRD) and the Debye-Scherrer equation \cite{Ingham:2015a},
the crystalline sizes extracted from Fig. \ref{Fig14} can be
evaluated as $69.8~$nm in a), $72.6~$nm in b), and $65.9~$nm in c).{\LARGE \
}

\section{Conclusion}

We have predicted theoretically and confirmed experimentally that the
localization of the optical field in the active three-dimensional disordered
percolating system has a nontrivial structure, allowing the generation and
propagation of localized field bunches. The system includes randomly
distributed nanoemitters, which, in general, corresponds to a spatially
averaged percolation environment. The localization strength and average
width of the corresponding structures have been analyzed, and that the
transition between the propagating bunches and point-like fields pinned to
the radiating emitters has been identified. The optical emission spectrum
from nanoemitters incorporated in the porous ceramic medium, measured in our
experiment, is in reasonable agreement with the prediction of the theory.

The analysis in this work is mainly concentrated on the emission pattern in
the frequency domain. As an extension of the work, we will study its
counterpart in the time domain, which requires the use of a more advanced
experimental technique.

\section*{Acknowledgments}

This work was supported, in part, by CONACYT (M\'{e}xico) through grant No.
A1-S-9201, and by the Israel Science Foundation through grant No. 1695/22.

\section*{Disclosures }

The authors declare no conflicts of interest.

\section*{Data Availability Statement}

Data that support the theoretical and experimental findings of this study
are available from the corresponding author upon a reasonable request.

\section*{Appendix: The synthesis of sol-gel nanocrystals at room
temperatures}

Ytterbium-doped zirconium nanocrystals were fabricated using the sol-gel
method. The produced nanocrystals feature different Yb percentages and were
characterized by EDS (energy dispersive spectrometry), XRD (X-ray
diffraction), and UV-Vis NIR (ultraviolet-visible near infrared)
spectroscopy. Note that in the theoretical part it was proposed to use pores
in the host ceramic medium to create a hybrid solid system with integrated
nanoemitters. However, in the experiment such a method is quite
time-consuming. It can be replaced by the following simple direct technique.
To create a hybrid mixture with incorporated nanoemitters, we apply the
well-known Sol-Gel technique \cite{sol-gel}, which makes it possible to make
hybrid medium directly by mixing the zirconium and the dopant. [I have
inserted here an obviously missing reference to the sol-gel technique.] %
For this purpose, a mixture of $1.0$
equivalent of zirconium n-propoxide ($70\%$) in absolute ethanol ($20$
equivalents) was stirred (\textquotedblleft agitated") for two minutes, and
then $0.1$ equivalents of HNO$_{3}$ ($67\%$) were drop-wise added under
intensive agitation. After that, $0.3$ equivalent of HCl ($37\%$) was added
under agitation too. A solution of $0.043$ equivalent of ytterbium
chloride-YbCl$_{3}$6H$_{2}$O ($99.99\%$) in $2$ mL of ethanol was added to
the mixture, agitated during $60$ min to produce the gel phase, and dried at
$100^{\mathrm{o}}$ C for 24 hours. The sample was annealed at $1000^{o}$ C
for 10 hours. The solid product was obtained in the form of fine white
particles with $1.23\%$ of ytterbium determined by EDS. When $0.087$
equivalents of YbCl$_{3}$6H$_{2}$O were{\LARGE \ }added, the obtained
product featured $2.36\%$ of ytterbium. A hydropneumatic press was used to
obtain the product in the form of thin ceramic pellets, which were used for
the absorption characterization.

\bibliographystyle{unsrt}
\bibliography{andercitBgnGMAb_r}

\end{document}